\documentclass[aps,twocolumn,letter,showpacs,preprintnumbers,amsmath,amssymb,superscriptaddress,bibnotes]{revtex4}

\usepackage{graphicx,epsfig}
\usepackage{dcolumn}
\usepackage{bm}

\begin{document}

\title{Missing Quasiparticles and the Chemical Potential Puzzle in the
Doping Evolution of the Cuprate Superconductors}

\author{K.M. Shen}
 \affiliation{Departments of Applied Physics, Physics, and Stanford
 Synchrotron Radiation Laboratory, Stanford University, Stanford,
 California 94305}

\author{F. Ronning}
 \altaffiliation[Present address: ]{Los Alamos National Laboratory, Los
 Alamos, NM 87545, USA}
 \affiliation{Departments of Applied Physics, Physics, and Stanford
 Synchrotron Radiation Laboratory, Stanford University, Stanford,
 California 94305}

\author{D.H. Lu}
 \affiliation{Departments of Applied Physics, Physics, and Stanford
 Synchrotron Radiation Laboratory, Stanford University, Stanford,
 California 94305}

\author{W.S. Lee}
 \affiliation{Departments of Applied Physics, Physics, and Stanford
 Synchrotron Radiation Laboratory, Stanford University, Stanford,
 California 94305}

\author{N.J.C. Ingle}
 \affiliation{Departments of Applied Physics, Physics, and Stanford
 Synchrotron Radiation Laboratory, Stanford University, Stanford,
 California 94305}

\author{W. Meevasana}
 \affiliation{Departments of Applied Physics, Physics, and Stanford
 Synchrotron Radiation Laboratory, Stanford University, Stanford,
 California 94305}

\author{F. Baumberger}
 \affiliation{Departments of Applied Physics, Physics, and Stanford
 Synchrotron Radiation Laboratory, Stanford University, Stanford,
 California 94305}

\author{A. Damascelli}
 \altaffiliation[Present address: ]{Department of Physics and Astronomy, University of British Columbia,
 Vancouver, B.C., Canada, V6T 1Z1}
 \affiliation{Departments of Applied Physics, Physics, and Stanford
 Synchrotron Radiation Laboratory, Stanford University, Stanford,
 California 94305}

\author{N.P. Armitage}
 \altaffiliation[Present address: ]{Department of Physics and
 Astronomy, University of California, Los Angeles, Los Angeles, CA 90095, USA}
 \affiliation{Departments of Applied Physics, Physics, and Stanford
 Synchrotron Radiation Laboratory, Stanford University, Stanford,
 California 94305}

\author{L.L. Miller}
 \affiliation{Department of Chemistry, University of Oregon, Eugene,
 Oregon 97403}

\author{Y. Kohsaka}
 \affiliation{Department of Advanced Materials Science,
 University of Tokyo, Kashiwa, Chiba 277-8651, Japan}

\author{M. Azuma}
 \affiliation{Institute for Chemical Research,
 Kyoto University, Uji, Kyoto 611-0011, Japan}

\author{M. Takano}
 \affiliation{Institute for Chemical Research,
 Kyoto University, Uji, Kyoto 611-0011, Japan}

\author{H. Takagi}
 \affiliation{Department of Advanced Materials Science,
 University of Tokyo, Kashiwa, Chiba 277-8651, Japan}

\author{Z.-X. Shen}
 \affiliation{Departments of Applied Physics, Physics, and Stanford
 Synchrotron Radiation Laboratory, Stanford University, Stanford,
 California 94305}

\date{\today}

\begin{abstract}

The evolution of Ca$_{2-x}$Na$_{x}$CuO$_{2}$Cl$_{2}$ from Mott
insulator to superconductor was studied using angle-resolved
photoemission spectroscopy. By measuring both the excitations near
the Fermi energy as well as non-bonding states, we tracked the
doping dependence of the electronic structure and the chemical
potential with unprecedented precision. Our work reveals failures
in the conventional quasiparticle theory, including the broad
lineshapes of the insulator and the apparently paradoxical shift
of the chemical potential within the Mott gap. To resolve this, we
develop a model where the quasiparticle is vanishingly small at
half filling and grows upon doping, allowing us to unify
properties such as the dispersion and Fermi wavevector with the
behavior of the chemical potential.

\end{abstract}



\maketitle

A central intellectual issue in the field of high-temperature
superconductivity is how an antiferromagnetic insulator evolves
into a superconductor. In principle, the ideal tool to address
this problem is angle-resolved photoemission spectroscopy (ARPES),
which can directly extract the single-particle excitations.
Despite the interest in this doping induced crossover, there
continues to be a lack of experimental consensus, perhaps the most
prominent example being the controversy over the chemical
potential, $\mu$. Over the past fifteen years, there have been
conflicting claims of $\mu$ either being pinned in mid-gap or
shifting to the valence / conduction band upon carrier doping
\cite{Damascelli03,Allen90,Shen91,Ino97,Steeneken03,Ronning03NaCCOC}.
The inability of photoemission spectroscopy to provide a logically
consistent understanding of this fundamental thermodynamic
quantity has been a dramatic shortcoming in the field.

In this paper, we present a new procedure to quantify $\mu$ with
unprecedented precision by ARPES, while allowing simultaneous high
resolution measurements on the low energy states. These
measurements allow us to make major conceptual advances in
addressing the doping evolution. We find that the long standing
confusion over $\mu$ stems from the manner in which
quasiparticle-like (QP) excitations in the doped samples emerge
from the unusually broad features in the undoped insulator. Our
work reveals inconsistencies in the conventional framework that
considers the main peak in the insulator spectrum to represent the
QP pole. On the one hand, we find that $\mu$ changes in a manner
consistent with an approximate rigid band shift; on the other
hand, this shift appears to occur \emph{within} the apparent Mott
gap of the parent insulator. We show that this ostensible paradox
can be naturally explained if one uses a model based on
Franck-Condon-like broadening (FCB) where the quasiparticle
residue, $Z$, is vanishingly small near half filling. This also
reconciles existing puzzles regarding the insulator and the
lightly doped compounds, and naturally ties the behavior of $\mu$
to low energy features such as the Fermi wavevector,
$\mathbf{k}_{\mathrm{F}}$, and the quasiparticle velocity
$v_{\mathrm{F}}$.

Ca$_{2-x}$Na$_{x}$CuO$_{2}$Cl$_{2}$ is an ideal system to address
the doping evolution of the cuprates. The stoichiometric parent
compound, Ca$_{2}$CuO$_{2}$Cl$_{2}$, is chemically stable and,
along with its close variants, has served as the prototype for the
undoped Mott insulator \cite{Damascelli03,Wells95}. Moreover, the
system possesses a simple structure, with only a single CuO$_{2}$
layer devoid of known superlattice modulations, structural
distortions, or surface states, unlike the Bi-based cuprates,
La$_{2-x}$Sr$_{x}$CuO$_{4}$, or YBa$_{2}$Cu$_{3}$O$_{7-\delta}$.
The \emph{x} = 0.10 and 0.12 samples had T$_{\mathrm{c}}$'s of 13
and 22 K, respectively (T$_{\mathrm{c,\,opt}}=28$ K), while the
\emph{x} = 0.05 composition was non-superconducting, and were
grown using a high pressure flux method \cite{KohsakaGrowth02}.
ARPES measurements were performed at Beamline \mbox{5-4} of the
Stanford Synchrotron Radiation Laboratory with typical energy and
angular resolutions of 13 meV and 0.3$^{\circ}$, respectively,
using photon energies of 21.2 and 25.5 eV. Measurements were
performed at 15 K, except for \emph{x} = 0, which was measured at
T $>$ 180 K.

\begin{figure}
\includegraphics{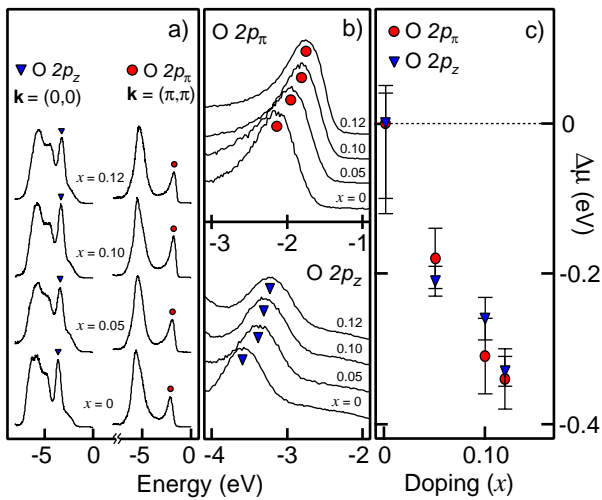} \vspace{0in}
\caption{(a) Valence band spectra for \emph{x} = 0, 0.05, 0.10,
and 0.12 compositions at $\mathbf{k} = (0,0)$ and $(\pi,\pi)$. O
$2p_{z}$ and O $2p_{\pi}$ states are marked by triangles and
circles, respectively. (b) Shifts of the O $2p_{z}$ and O
$2p_{\pi}$ peaks shown on an expanded scale. (c) Doping dependence
of $\mu$.} \label{ValenceBand}
\end{figure}

Previous quantitative studies of $\mu$ have relied on core level
spectroscopy. However, the precision of this method is limited and
its interpretation can be very complicated. Since measuring $\mu$
is of paramount importance, we introduce a new approach which we
believe to be more accurate and direct, and can be performed in
parallel with ARPES on the near-E$_{\mathrm{F}}$ states. Our
method utilizes delocalized, non-bonding O $2p$ states in the
valence band, in particular, O $2p_{z}$ at (0,0) and O $2p_{\pi}$
at $(\pi,\pi)$ shown in Figure \ref{ValenceBand}a. These were
identified in earlier works and have no overlap with the Cu
$3d_{x^2 - y^2}$ orbital or Zhang-Rice singlet
\cite{Pothuizen97,Hayn99}. Because these well-defined peaks are
relatively close to E$_{\mathrm{F}}$ and measured at a single
wavevector, we can treat these states as delocalized bands. The
overall shape of the valence band remains unchanged with doping,
suggesting a rigid band shift on a gross scale ($\sim 10$ eV). In
Figures \ref{ValenceBand}b and \ref{ValenceBand}c, we show the
shift of the O $2p_{z}$ and O $2p_{\pi}$ peaks on an expanded
scale with statistics collected from $> 5$ samples for each
concentration. All data are referenced to the \emph{x} = 0
composition, and we describe the methodology for determining
$\mu_{0}$ later in the text. The shift from these marker states
yields $\mu_{0.05} = -0.20$ eV, $\mu_{0.10} = -0.28$ eV, and
$\mu_{0.12} = -0.33$ eV, all relative to $\mu_{0}$, with a typical
uncertainty of $\pm 0.025$ eV. At finite \emph{x}, $\frac{\partial
\mu}{\partial x} = 1.8 \pm 0.5$ eV / hole, comparable to band
structure ($\sim$ 1.3 eV / hole) \cite{Mattheiss90}.

\begin{figure}[b!]
\includegraphics{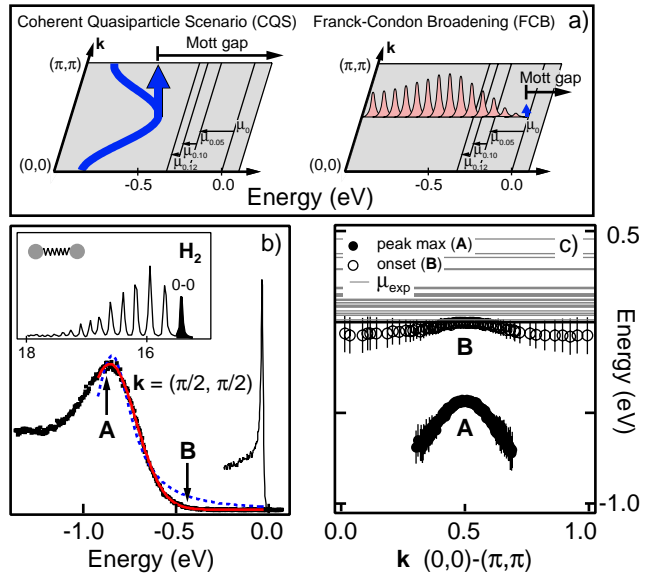} \vspace{0in}
\caption{(a) Illustrations of the Coherent Quasiparticle Scenario
(CQS) and the Franck-Condon Broadening (FCB) model. (b)
Ca$_{2}$CuO$_{2}$Cl$_{2}$ at $\mathbf{k} = (\pi/2,\pi/2)$ with
fits to a spectral function (dashed blue) and gaussian (red). A
and B denote the peak maximum and the onset of spectral weight,
respectively. Comparison with Sr$_{2}$RuO$_{4}$ is shown (thin
black). Upper inset shows photoemission spectra from H$_2$, after
Ref. \onlinecite{Turner70}. (c) Dispersion of A and B along
$(0,0)-(\pi,\pi)$, along with experimental values for $\mu$
(lines).} \label{Insulator}
\end{figure}

These measurements reveal a fundamental failure of the traditional
framework where the main peak of the insulator represents a
quasiparticle pole, which we call the ``coherent quasiparticle
scenario'' (CQS). Here, all energies above the peak maximum should
fall within the Mott gap. As shown in Figure \ref{ValenceBand},
$\mu$ shifts by an amount compatible with predictions from band
structure calculations. However, this shift appears to occur
\emph{within the apparent Mott gap} of the parent insulator - a
logical inconsistency as there are no available states within the
gap to shift into, as illustrated at the left of Figure
\ref{Insulator}a. While impurity-like states may form within the
gap, in this picture $\mu$ should not drop so rapidly with doping.
In the remainder of this paper, we combine the measurements of
$\mu$ with high resolution studies of the near-E$_{\mathrm{F}}$
states. Our results suggest a model which borrows heavily from the
concept of FCB, shown at the right of Figure \ref{Insulator}a.

To understand the failings of the CQS, the obvious starting point
is the parent antiferromagnetic insulator. Early studies of
Ca$_{2}$CuO$_{2}$Cl$_{2}$ and Sr$_{2}$CuO$_{2}$Cl$_{2}$ yielded
broad peaks which exhibited a dispersion consistent with
calculations for the extended $t-J$ model
\cite{Wells95,Tohyama00,Damascelli03}. These peaks were
effectively interpreted as quasiparticle poles, in the context of
the CQS. However, one crucial point that remained unresolved was
the extreme width of these excitations. We address this as a
critical weakness of the CQS, and use this as a starting point for
constructing a new model. Understanding this lineshape is crucial,
since many celebrated features such as the $d$-wave gap, the
superconducting peak, dispersion anomalies, and the pseudogap
\cite{Damascelli03}, necessarily emerge from this starting point.
Data taken at the top of the lower Hubbard band, $\mathbf{k} =
(\pi/2,\pi/2)$, is shown in Figure \ref{Insulator}b. In the CQS,
one would expect the peak width, $\Gamma$, to be extremely narrow
at the top of the band due to phase space constraints, analogous
to excitations at the Fermi energy in a metal. Instead, $\Gamma$
is comparable to the entire bandwidth $2J \sim 350$ meV,
completely inconsistent with such a picture. Moreover, the width
cannot be due to disorder, as the undoped system is
stoichiometric, and adding chemical dopants results in sharper
structures, as will be shown. For comparison, we also present
spectra from Sr$_{2}$RuO$_{4}$ (thin black) exhibiting a nearly
resolution-limited peak. Given that well-defined QP excitations
can be observed by ARPES, we must confront the origin of the broad
peaks in Ca$_{2}$CuO$_{2}$Cl$_{2}$. Moreover, in the CQS, the peak
in Figure \ref{Insulator}b should be well described by a spectral
function $\mathcal{A}(\mathbf{k},\omega) = -\frac{1}{\pi}
\frac{\Sigma^{\prime\prime}}{(\omega - \epsilon_{\mathbf{k}} -
\Sigma^{\prime})^2 + (\Sigma^{\prime\prime})^2}$, which should be
approximately Lorentzian with a width dominated by an impurity
scattering term, $\Gamma_{\mathrm{imp}}$ \cite{Damascelli03}. A
fit of $\mathcal{A}(\mathbf{k},\omega)$ to the experimental data
is shown and agrees poorly; to achieve even this,
$\Gamma_{\mathrm{imp}}$ was assumed to be unphysically large
($\sim$ 300 meV), given that the material is stoichiometric and
free of chemical dopants.

In light of this failure of the CQS, we believe that an analogy to
one of the simplest quantum systems, the H$_{2}$ molecule, may be
enlightening. The H$_{2} \rightarrow$ H$_{2}^{+}$ photoemission
spectrum, shown in the upper inset of Figure \ref{Insulator}b,
exhibits FCB. Only the `0-0' peak (filled black) represents the
H$_{2}^{+}$ final state with no excited vibrations and comprises
only $\sim$ 10\% of the intensity, while transitions to excited
states with $n$ = 1, 2, 3, and 4 vibrational quanta possess higher
intensities than 0-0. In the solid state, 0-0 alone would
represent the QP or the coherent part of the spectral function,
$\mathcal{A}_{\mathrm{coh}}$, whereas the excited states comprise
the incoherent part, $\mathcal{A}_{\mathrm{inc}}$. This behavior
is redolent of polarons, and such models have been suggested in
systems where strong couplings are present
\cite{Dessau98,Perebeinos00,Perfetti02}. The low energy tail is
suppressed exponentially, inconsistent with power law falloffs
from $\mathcal{A}(\mathbf{k},\omega)$ spectral functions, but
well-described by a Gaussian FCB envelope.

Another unresolved issue is a large energy scale separating the
peak from the experimental positions of $\mu$ inside the Mott gap.
For an insulator, $\mu$ is not well defined, and is pinned by
surface defects and impurities and will vary between samples.
However, the limits of this distribution are well defined, with a
lower bound set by the QP pole at the top of the valence band. For
this study, we identify two features : the peak maximum (A), and
the onset of intensity (B), where B is determined from the first
statistically significant signal above background ($3\sigma$). In
Figure \ref{Insulator}c, we show the dispersion of A and B along
(0,0)-($\pi,\pi$). While A qualitatively tracks the dispersion of
the $t-J$ model, B disperses only weakly and has a large
separation of $> 450$ meV from A. We present the distribution of
$\mu_{\mathrm{exp}}$ from a large number of samples in Figure
\ref{Insulator}c and B clearly sets a lower bound for the
distribution of $\mu_{\mathrm{exp}}$. This behavior suggests FCB
where the true QP (B) is hidden within the tail of spectral
intensity, and A is simply incoherent weight associated with
shake-off excitations. For \emph{x} = 0, we reference A and the
valence band features such that B is aligned to 0, and this is
consistent with the lowest measured value for
$\mu_{\mathrm{exp}}$. This demarcates the upper bound for both the
QP at half filling and $\mu(x = 0^{+})$, not the actual position
of $\mu$ for \emph{x} = 0, and provides us with the value for
$\mu_{0}$ which we have used throughout the paper. This model is
also consistent with the temperature dependence of the lineshape,
where a similar multiple initial/final state model was proposed
\cite{Kim02}. At this stage, we cannot distinguish which
interactions are causing this broadening. Although numerical
simulations of the $t-J$ model predict that $Z$ remains finite ($Z
\sim 0.2$) \cite{Dagotto94}, some analytical calculations have
predicted that interactions with the antiferromagnetic background
cause $Z \rightarrow 0$ \cite{Shraiman88,Sheng96}. Another
possibility is the coupling to the lattice, and our data bears
some resemblance to that of lattice polaronic systems (1D Peierls,
manganites) \cite{Perfetti02,Dessau98,Perebeinos00}. We note that
a recent calculation incorporating lattice effects in the $t-J$
model has closely reproduced the ARPES spectra, including a
vanishing QP peak (B) and a broad hump (A) which recovers the
original $t-J$ dispersion \cite{Mishchenko04}.

\begin{figure}[t!]
\includegraphics{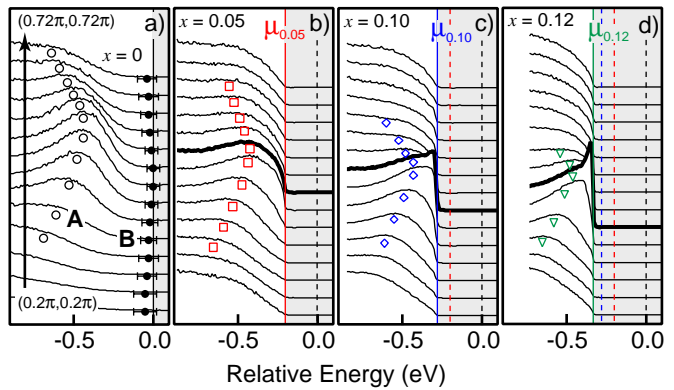} \vspace{0in}
\caption{(a-d) EDC spectra from \emph{x} = 0, 0.05, 0.10, and 0.12
from $(0.2\pi,0.2\pi)$ to $(0.72\pi,0.72\pi)$ with hump positions
marked by open symbols and $\mathbf{k}_{\mathrm{F}}$ shown in
bold. Data are plotted on a relative energy scale referenced to
the shift in $\mu$ shown in Figure \ref{ValenceBand}c.}
\label{EDCStacks}
\end{figure}

\begin{figure}
\includegraphics{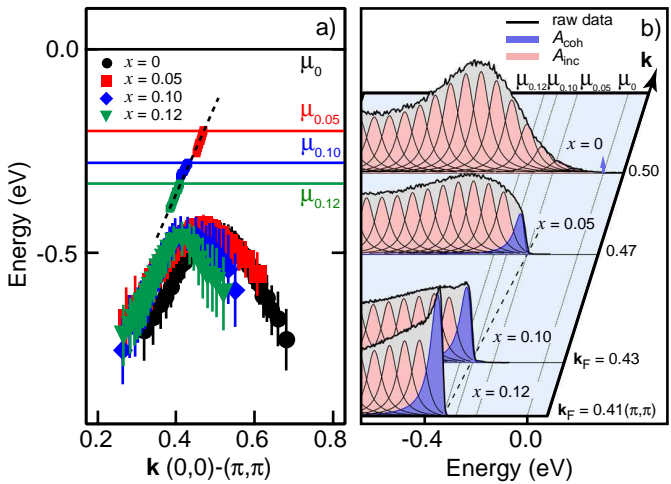} \vspace{0in}
\caption{(a) Summary of hump (symbols) and MDC dispersions (lines)
from Figure \ref{EDCStacks}. (b) Doping dependence of spectra from
$\mathbf{k}_{\mathrm{F}}$ along with a schematic of the proposed
distribution of coherent (blue) and incoherent (pink) spectral
weight.} \label{ShiftCartoon}
\end{figure}

In Figure \ref{EDCStacks}, we show the doping evolution of the
near-E$_{\mathrm{F}}$ energy distribution curves (EDCs), from
($0.2\pi,0.2\pi$) to ($0.72\pi,0.72\pi$). All data are plotted on
a fixed energy scale relative to $\mu_{0}$ using the values
determined in Figure \ref{ValenceBand}c. With doping, feature A
evolves smoothly into a broad, high energy hump with a backfolded
dispersion that qualitatively reflects the parent insulator
(symbols), while $\mu$ shifts from B into the lower Hubbard band.
It is now clear that $\mu$ does not fall immediately to A upon
hole doping as expected in the CQS. Spectral weight develops at
$\mu$, and a well-defined peak becomes visible for the \emph{x} =
0.10 and 0.12 compositions, comprising a coherent, low-energy
band. The dispersion of the hump is summarized in Figure
\ref{ShiftCartoon}a, and was determined by tracking local maxima
in the EDCs; where the hump becomes less distinct, we also use the
second derivative of the EDCs. We note that in our model, A no
longer represents any precise physical quantity. We also track the
dispersion of the lowest energy excitations (-0.05 eV $ < \omega
<$ E$_{\mathrm{F}}$) from a momentum distribution curve (MDC)
analysis (lines). The collective behavior of the data reveals that
A is roughly fixed at high energies (-450 meV) with doping,
justifying our FCB model which, in some sense, decouples A from
$\mu$. The dispersion of the low-energy states reveals a
remarkable universal behavior across doping levels where both the
velocities of the QP dispersion ($v_{\mathrm{F}}$) and Fermi
wavevectors ($\mathbf{k}_{\mathrm{F}}$) virtually collapse onto a
single straight line with a band velocity (1.8 eV$\cdot$\AA)
corresponding closely to the recently discovered ``universal nodal
velocity'' \cite{Zhou03}. This result also ties the chemical
potential to the QP dispersion to naturally explain how
$\mathbf{k}_{\mathrm{F}}$ evolves with doping, by simply sliding
$\mu$ down the true QP band and calculating
$\Delta\mathbf{k}_{\mathrm{F}} \sim \Delta\mu / v_{\mathrm{F}}$,
as shown in Figure \ref{ShiftCartoon}a. We note that a number of
different theoretical proposals have predicted the emergence of
sharp QP-like excitations from broad features, including dynamical
mean-field theories \cite{Georges96} or ``gossamer''
superconductivity \cite{Bernevig03}.

A schematic cartoon overlaid on experimental data is shown in
Figure \ref{ShiftCartoon}b, illustrating the doping dependence of
$\mu$ and the proposed transfer between coherent (blue) and
incoherent (pink) spectral weight. This also explains the lack of
any well-defined peak at $\mathbf{k}_{\mathrm{F}}$ for \emph{x} =
0.05, expected in Fermi liquid models, which can be naturally
explained by the vast incoherent weight overwhelming any small
coherent peak. This also clarifies whether in-gap states or a
rigid band shift was the correct description of the doping
evolution, and it is now evident that neither are adequate in the
FCB context. A similar emergence of QP excitations at low dopings
was also observed in La$_{2-x}$Sr$_{x}$CuO$_{4}$ \cite{Yoshida03},
although one distinction between
Ca$_{2-x}$Na$_{x}$CuO$_{2}$Cl$_{2}$ and
La$_{2-x}$Sr$_{x}$CuO$_{4}$ is the relative separation between A
and E${_\mathrm{F}}$ \cite{Damascelli03}.

In conclusion, we have developed a phenomenological model based on
high precision measurements of $\mu$ and detailed studies of the
near-E$_{\mathrm{F}}$ states, providing us with the first globally
consistent understanding of the doping evolution of the cuprates.
This picture can be summarized as follows : i) At half filling,
$Z$ is vanishingly small in a manner reminiscent of Franck-Condon
broadening. The true QP is found in the tail of spectral
intensity, approximately 450 meV above the peak position. ii) The
previous misidentification of the peak maximum as the QP pole was
at the root of the long standing confusion over $\mu$. iii) With
doping, spectral weight is transferred to the low-energy QP-like
peak. iv) The shift of the chemical potential and
$\mathbf{k}_{\mathrm{F}}$ is dictated by the band velocity of this
faint QP band. We believe that the above picture provides a
foundation for the origin of the quasiparticles upon doping and
should be used as a guide to develop microscopic theories for
high-T$_{\mathrm{c}}$ superconductivity.

We would like to thank A. Fujimori and C. Kim for enlightening
discussions. SSRL is operated by the DOE Office of Basic Energy
Science under contract DE-AC03-765F00515. K.M.S. acknowledges SGF
and NSERC for their support. The ARPES measurements at Stanford
were also supported by NSF DMR-0304981 and ONR N00014-98-1-0195.

\end{document}